\documentclass[aps,pra,twocolumn,showpacs,amsmath,amssymb,groupedaddress]{revtex4-1}
\usepackage{graphicx}

\begin{document}
\title{Bose-Einstein condensates in the presence of Weyl spin-orbit coupling}
\author{Ting Wu and Renyuan Liao}
\affiliation{Fujian Provincial Key Laboratory for Quantum Manipulation and New Energy Materials, College of Physics and Energy, Fujian Normal University, Fuzhou 350108, China}
\affiliation{Fujian Provincial Collaborative Innovation Center for Optoelectronic Semiconductors and Efficient Devices, Xiamen, 361005, China}
\date{\today}
\begin{abstract}
 We consider two-component Bose-Einstein condensates subject to Weyl spin-orbit coupling. We obtain mean-field ground state phase diagram by variational method. In the regime where interspecies coupling is larger than intraspecies coupling, the system is found to be fully polarized and condensed at a finite momentum lying along the quantization axis. We characterize this phase by studying  the excitation spectrum, the sound velocity, the quantum depletion of condensates, the shift of ground state energy, and the static structure factor. We find that spin-orbit coupling and interspecies coupling generally leads to competing effects.
\end{abstract}
\maketitle
\section{Introduction}
The creation of synthetic gauge fields in ultracold atomic gases provides fascinating opportunities for exploring quantum many-body physics~\cite{DAL11}. Of particular interest is the realization of non-Abelian spin-orbit coupling (SOC)~\cite{GAL13,GOL14,ZHA15}. Spin-orbit coupling is crucial for realizing intriguing phenomena such as the quantum spin Hall effect~\cite{JAI15}, new materials classes such as topological insulators and superconductors~\cite{HAN10,QI11,BER13}. In bosonic systems, the presence of SOC may lead to novel ground states that have no known analogs in conventional solid-state materials~\cite{STA08,ZHA10,HO11}. In cold atomic gases, spin-orbit coupling can be implemented by Raman dressing of atomic hyperfine states~\cite{LIN11,PAN12}. The tunability of the Raman coupling parameters promises a highly flexible experimental platform to explore interesting physics resulting from spin-orbit coupling~\cite{SPI15}. Recently, two-dimensional SOC has been experimentally realized in cold atomic gases~\cite{HUA15,PAN16}.

In anticipation of immediate experimental relevance, intense theoretical attention has been paid to the physics of ultracold atomic gases in the presence of SOC~\cite{GOL14,ZHA15}.  In the absence of interparticle interactions, the low-lying density of states is two-dimensional for Rashba-type SOC~\cite{STA08}. In particular, the single-particle energy minimum featured a Rashba-ring, which has important consequences on the ground state and finite-temperature properties of SOC Bose gases~\cite{SAN11,HU12,KAW12,UED12,BAR12,BAY12,CUI13,QI13,YUN13,LIA14}, as the role of quantum fluctuations gets enhanced due to huge degeneracies at the lowest-lying states. The three-dimensional analog of Rashba-type SOC is interesting because it is expected to stabilize a long-sought skyrmion mode in the ground state of trapped Bose-Einstein condensates (BECS)~\cite{KAW12,ZHOU13,WU16}. This Weyl-type SOC can be implemented following the proposals~\cite{AND12,WU12,UED13} by using powerful quantum technology. Although there is currently no evidence for Weyl fermions to exist as fundamental particles in our universe, Weyl-like quasiparticles have been detected recently in condensed-matter systems~\cite{HAS15,DIN15}. In light of these discoveries, the study of Weyl SOC in ultracold atom systems becomes particularly relevant, since the ability of manipulate the Weyl-SOC strength creates interesting opportunities for the exploration of effects not predicted in the realm of particle physics. In addition, the study of the effects of SOC may reveal some interesting physics unexplored in conventional binary Bose condensates~\cite{LAW97,HAL98}. In this work, we shall examine the physics of two-component Bose gases subject to Weyl-type SOC. Firstly, we will introduce the model and determine the mean-field ground state by variation approach. Secondly, we will set out to study a particular realization of ground state where quantum fluctuation plays an essential role. Specially, we will investigate the interplay of spin-orbit coupling and interspecies interaction upon the ground state properties of the system. Finally, we will come to a summary.

\section{Model and Formalism}
We consider a 3D homogeneous interacting two-component Bose gas subject to Weyl-type spin-orbit coupling, described by the Hamiltonian $H=H_0+H_I$, with
\begin{subequations}
\begin{eqnarray}
     H_0&=&\int d^3\mathbf{r} \Psi^\dagger(\mathbf{r})\left[-\frac{\hbar^2\nabla^2}{2m}+\lambda\vec{\mathbf{\sigma}}\cdot\hat{\mathbf{p}}\right]\Psi(\mathbf{r}),\\
     H_I&=&\int d^3\mathbf{r}\left[g\sum_\sigma n_\sigma(\mathbf{r})^2+2g_{\uparrow\downarrow}n_{\uparrow}n_{\downarrow}\right].
\end{eqnarray}
\end{subequations}
Here $\Psi(\mathbf{r})=(\psi_\uparrow,\psi_\downarrow)^T$ is a two-component spinor field, $\vec{\sigma}=\hat{x}\sigma_x+\hat{y}\sigma_y+\hat{z}\sigma_z$, $\hat{\mathbf{p}}$ is the momentum operator, $n_
\sigma=\psi_\sigma^\dagger\psi_\sigma$ is the density for component $\sigma\in\{\uparrow,\downarrow\}$, $\lambda$ is the strength of the spin-orbit coupling, and the strength for the intraspecies interaction and interspecies interaction is $g$ and $g_{\uparrow\downarrow}$, respectively. For brevity, we set $\hbar=2m=1$ from now on.

Diagonalization of $H_0$ yields the two-branch single-particle energy spectrum $E_\pm(\mathbf{p})=p^2\pm\lambda p$, and the corresponding eigenfunctions are given by
\begin{eqnarray}
    \Phi_\pm(\mathbf{p})=\begin{pmatrix} \sin{\left[(\pi-2\theta_\mathbf{p}\pm\pi)/4\right]}e^{-i\varphi_\mathbf{p}}\\
    \cos{\left[(\pi-2\theta_\mathbf{p}\pm\pi)/4\right]}
    \end{pmatrix}\frac{e^{i\mathbf{p}\cdot\mathbf{r}}}{\sqrt{V}},
\end{eqnarray}
where $V$ is the volume of the system. The lowest-energy state for a given propagating direction parameterized by $\theta_\mathbf{p}$ and $\varphi_\mathbf{p}$ is from the ``-" branch and occurs at momentum $\mathbf{p}=\frac{\lambda}{2}(\sin\theta_\mathbf{p}\cos{\varphi_\mathbf{p}},\sin\theta_\mathbf{p}\sin{\varphi_\mathbf{p}},\cos\theta_\mathbf{p})$.

\begin{figure}[t]
\includegraphics[width=1\columnwidth]{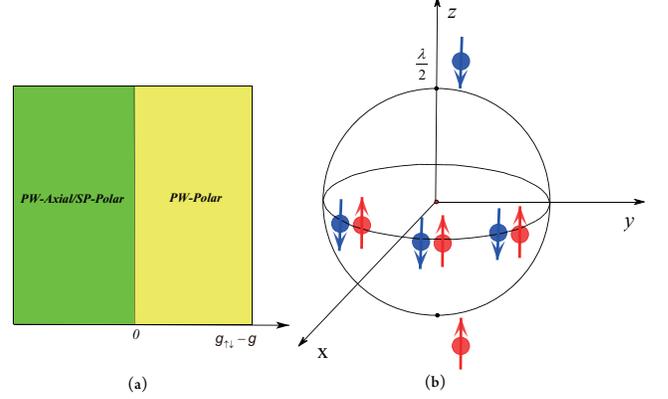}
\caption{(color online) Mean-field ground state phase diagram. Panel (a): for $g_{\uparrow\downarrow}-g>0$, the ground state is in PW-Polar phase, where it is a plane wave with the condensation momentum being parallel to the $z$-axis. For $g_{\uparrow\downarrow}-g<0$, the system may be in the phase of either PW-Axial or SP-Polar. Here PW-Axial phase stands for one plane wave with the condensation momentum lying in the $x$-$y$ plane, and SP-Polar phase stands for the condensation at two opposite momenta along the $z$-axis. Panel (b): schematic representation of the PW-Polar and PW-Axial phases. For one plane-wave condensation at either the north pole or the south pole is called the PW-Polar phase, it is a fully polarized phase as only one component is allowed. For one plane-wave condensation in the $x$-$y$ plane is called PW-Axial, it is an unpolarized phase.
}
\label{fig1}
\end{figure}

To determine the ground state of an interacting system, as routinely done in the literature~\cite{ZHA10,YUN12,CUI13,SUN15,LIA15}, we assume that the system has condensed into a coherent superposition of two plane-wave states with opposite momenta with magnitude $p=\lambda/2$. Thus the condensate wave function adopts the form  $\Phi_0=C_+\Phi_-(\mathbf{p})+C_-\Phi_-(\mathbf{-p})$, where $C_+$ and $C_-$ are two complex numbers to be determined and subject to normalization condition $|C_+|^2+|C_-|^2=n_0$. Without loss of generality, the normalization condition suggests the parametrization $|C_+|^2=n_0\cos^2{(\alpha/2)}$ and $|C_-|^2=n_0\sin^2{(\alpha/2)}$, with $\alpha\in[0,\pi]$. Upon substitution into $E_G=<\Phi_0|H|\Phi_0>$, the variational ground state energy per particle is evaluated as
\begin{eqnarray}
&&\frac{E_G}{n_0V}=-\frac{\lambda^2}{4}+gn_0+\frac{(g_{\uparrow\downarrow}-g)n_0}{2}f(\theta_\mathbf{p},\alpha),
\end{eqnarray}
where $f(\theta_\mathbf{p},\alpha)=\sin^2{\theta_\mathbf{p}}+\sin^2\alpha-3\sin^2{\theta_\mathbf{p}}\sin^2\alpha/2$. Minimization of the ground state energy with respect to $\theta_\mathbf{p}$ and $\alpha$, one obtains the ground state phase diagram, summarized in Fig.$\ref{fig1}$. When $g_{\uparrow\downarrow}-g>0$, the system is found to be in the phase of PW-Polar, which is a fully polarized phase with condensation momentum lying along the quantization axis. When $g_{\uparrow\downarrow}-g<0$, at mean-field level, there are two degenerate phases: one is unpolarized PW-Axial phase, which is condensed at one plane-wave with momentum lying in the $x$-$y$ plane; the other one is the SP-Polar phase, which is striped phase mixing of two opposite momentum along the $z$-axis. There exists a critical point when $g_{\uparrow\downarrow}-g=0$. In this case the system enjoys a SU(2) pseudo-spin rotation symmetry. To determine which phase the system prefers requires calculation going beyond mean field, and in principle it is believed to lead to a unique ground state via the mechanism of ``order from disorder''~\cite{HAN12,HU12}.

Within the imaginary-time field integral, the partition function of the system may be cast as~\cite{SIM06} $\mathcal{Z}=\int \mathcal{D}[\psi_\sigma^*\psi_\sigma]e^{-S[\psi_\sigma^*,\psi_\sigma]}$, with the action
$S=\int_0^\beta d\tau[d\mathbf{r}\sum_\sigma\psi_\sigma^*(\partial_\tau-\mu)\psi_\sigma+H(\psi_\sigma^*,\psi_\sigma)]$, where $\beta=1/T$ is the inverse temperature and $\mu$ is the chemical potential introduced to fix the total particle number. Here, for simplicity, we restrict ourself to studying the PW-Polar phase. Without loss of generality, we further assume that the condensation occurs at momentum $\vec{\kappa}=(0,0,-\lambda/2)$, then the ground state wave function is determined as $\Phi_0=\sqrt{n_0}(1,0)^Te^{-i\lambda z/2}$. It is a fully polarized phase with condensation momentum aligning antiparallel with the quantization axis. We split the Bose field into the mean-field part $\phi_{0\sigma}$ and the fluctuating part $\phi_{\mathbf{q}\sigma}$ as $\psi_{\mathbf{q}\sigma}=\phi_{0\sigma}\delta_{\mathbf{q}\vec{\kappa}}+\phi_{\mathbf{q}\sigma}$. After substitution, the action can be formally written as $S=S_0+S_f$, where $S_0=\beta V\left[(-\frac{\lambda^2}{4}-\mu)n_0+gn_0^2\right]$ is the mean-field contribution and $S_f$ denotes a contribution from the fluctuating fields. The chemical potential may be determined via saddle point condition $\partial S_0/\partial n_0=0$, yielding $\mu=-\frac{\lambda^2}{4}+2gn_0$. At this point, the action is exact. However, it contains terms of cubic and quartic orders in fluctuating fields. To proceed, we resort to the celebrated Bogoliubov approximation, where only terms of zeroth and quadratic orders in the fluctuating fields are retained. By defining a four-dimensional column vector $\Phi_{\mathbf{q}}=(\phi_{\vec{\kappa}+\mathbf{q}\uparrow},\phi_{\vec{\kappa}+\mathbf{q}\downarrow},\phi_{\vec{\kappa}-\mathbf{q}\uparrow}^*,\phi_{\vec{\kappa}-\mathbf{q}\downarrow}^*)$, we can bring the fluctuating part of the action into the compact form $S_f\approx\sum_{\mathbf{q},iw_n}  \frac{1}{2}\Phi_{\mathbf{q}}^\dagger\mathcal{G}^{-1}(\mathbf{q},iw_n)\Phi_\mathbf{q}-\frac{\beta}{2}\sum_{\mathbf{q},\sigma}\epsilon_{\mathbf{q}\sigma}$,where $w_n=2\pi n/\beta$ is the bosonic Matsubar frequencies, and the inverse Green's function $\mathcal{G}^{-1}(\mathbf{q},iw_n)$ is defined as

\begin{widetext}
\begin{eqnarray}
\mathcal{G}^{-1}=\begin{pmatrix}
 -iw_n+\epsilon_\mathbf{q\uparrow} &  R_\mathbf{q}  &  2gn_0 & 0  \\
 R_\mathbf{q}^*  &  -iw_n+\epsilon_\mathbf{q\downarrow} & 0 & 0\\
 2gn_0 & 0 &  iw_n+\epsilon_{-\mathbf{q}\uparrow}& R_{-\mathbf{q}}^*\\
 0 & 0&  R_{-\mathbf{q}}& iw_n+\epsilon_{-\mathbf{q}\downarrow}
\end{pmatrix},
\end{eqnarray}
\end{widetext}
where  $\epsilon_{\mathbf{q}\uparrow}=q^2+2gn_0$, $\epsilon_{\mathbf{q}\downarrow}=q^2-2\lambda q_z+\lambda^2+2(g_{\uparrow\downarrow}-g)n_0$, and $R_\mathbf{q}=\lambda(q_x-iq_y)$. Throughout our calculation, we will choose $gn_0$ as a basic energy scale and $\sqrt{gn_0}$ as the corresponding momentum scale. To characterize the strength of interspecies coupling, we define a dimensionless parameter $\eta=g_{\uparrow\downarrow}/g$.
\section{Calculation and Results}
\begin{figure}[t]
\includegraphics[width=1.0\columnwidth]{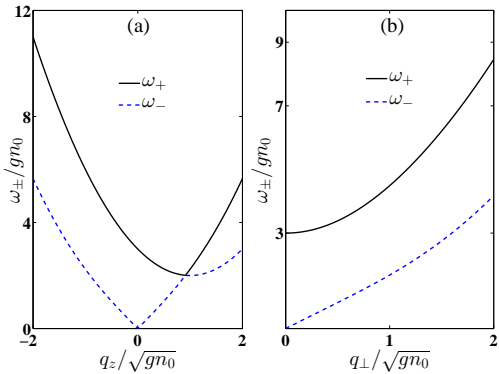}
\caption{(color online) Two branches of excitation spectrum $\omega_{\pm}$ in the momentum space: (a) along the $z$-axis, and (b) in the $x$-$y$ plane. Here we set interspecies coupling $\eta=2.0$ and spin-orbit coupling $\lambda=\sqrt{gn_0}$.
}
\label{fig2}
\end{figure}

The excitation spectrum of the system can be found by examining the poles of the Green's function $\mathcal{G}(\mathbf{q},iw_n)$. To achieve this , one proceeds by evaluating the determinant of $\mathcal{G}^{-1}(\mathbf{q},iw_n)$,
  \begin{eqnarray}
  det[\mathcal{G}^{-1}]&=&(iw_n^2-\omega_{10}^2)\left[(iw_n+2\lambda q_z)^2-\omega_{20}^2\right]-2\lambda^2q_\perp^2F,\nonumber\\
  F&=&iw_n(iw_n+2\lambda q_z)+(q^2+2gn_0)\omega_{20}-\frac{\lambda^2q_\perp^2}{2},\nonumber\\
  \label{eq:4}
\end{eqnarray}
where $\omega_{10}=q\sqrt{q^2+4gn_0}$ and $\omega_{20}=\lambda^2+q^2+2(g_{\uparrow\downarrow}-g)n_0$, and $q_\perp=q_x^2+q_y^2$. By solving the secular equation $DetG^{-1}(\mathbf{q},\omega_{\mathbf{q}s})=0$, one finds two branches of excitation spectrum $\omega_{\mathbf{q}\pm}$. As seen from Eq.~($\ref{eq:4})$, the excitation spectrum enjoys the azimuthal symmetry. Therefore we only plot the spectrum along two typical directions in Fig.~$\ref{fig2}$. Along the $z$-axis, the lower branch show the features of roton-maxon structure, indication of the tendency toward crystallization~\cite{YUN12}. Such roton-maxon spectrum has been detected in recent Bragg spectroscopy experiments~\cite{CHI15,PAN15,KHA14}, and the spectrum is asymmetrical with respect to reversing the direction. In the $x$-$y$ plane, the two branches are well separated as the upper branch is gapped while the lower branch becomes gapless as it approaches the origin $\mathbf{q}=(0,0,0)$.

\begin{figure}[t]
\includegraphics[width=1.0\columnwidth]{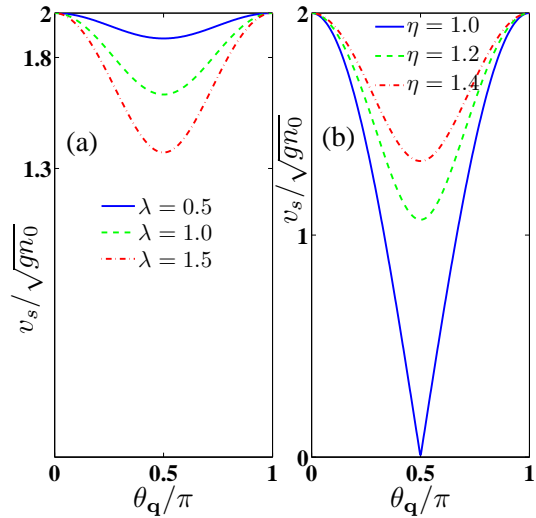}
\caption{(color online) Polar angle $\theta_\mathbf{q}$ dependence of the sound velocity $v_s$: (a) for different spin-orbit coupling strength $\lambda$ at $\eta=2.0$; (b) for different interspecies coupling $\eta$ at $\lambda=\sqrt{gn_0}$.
}
\label{fig3}
\end{figure}

Aside from the roton mode discussed above, the lower branch of the excitation spectrum also contains important information about the photon mode.  Along $z$ direction where $q_\perp=0$, it is straight forward to analytically derive two branches of solutions from Eq.~($\ref{eq:4}$): $\omega_-=\omega_{10}$ and $\omega_+=-2\lambda q_z+\omega_{20}$. The sound velocity along this direction is $v_z^s=2\sqrt{gn_0}$. In the $x$-$y$ plane, low-energy expansion around the gapless point $(0,0,0)$ yields $\omega_-\approx v_\perp^s q_\perp+\mathcal{O}(q_\perp^2)$ with in-plane isotropic sound velocity given by  $v_\perp^s=\sqrt{2gn_0}\sqrt{2(\eta-1)/[2(\eta-1)+\lambda^2/gn_0]}$. Numerically we compute the sound velocity via $v_s(\mathbf{q})=\lim_{q\rightarrow 0}\omega_-(\mathbf{q})/q$. We find that the sound velocity varies with the polar angle $\theta_\mathbf{q}$, as shown in Fig.~$\ref{fig3}$. The sound velocity enjoys a symmetry of $v_s(\theta_\mathbf{q})=v_s(\pi-\theta_\mathbf{q})$, with the maximum sound velocity achieved along $z$-axis and the minimum one in the $x$-$y$ plane. Away from the critical point where $\eta=1$, the spin-orbit coupling suppresses the sound velocity along any polar direction except for $\theta_\mathbf{q}=0$ and $\pi$, as indicated in Fig.~$\ref{fig3}(a)$. Interestingly, as seen in Fig.~$\ref{fig3}(b)$, suppression of sound velocity due to spin-orbit coupling could be mitigated by increasing the interspecies coupling, an indication of competing effects of spin-orbit coupling and interspecies coupling.

\begin{figure}[t]
\includegraphics[width=1.0\columnwidth]{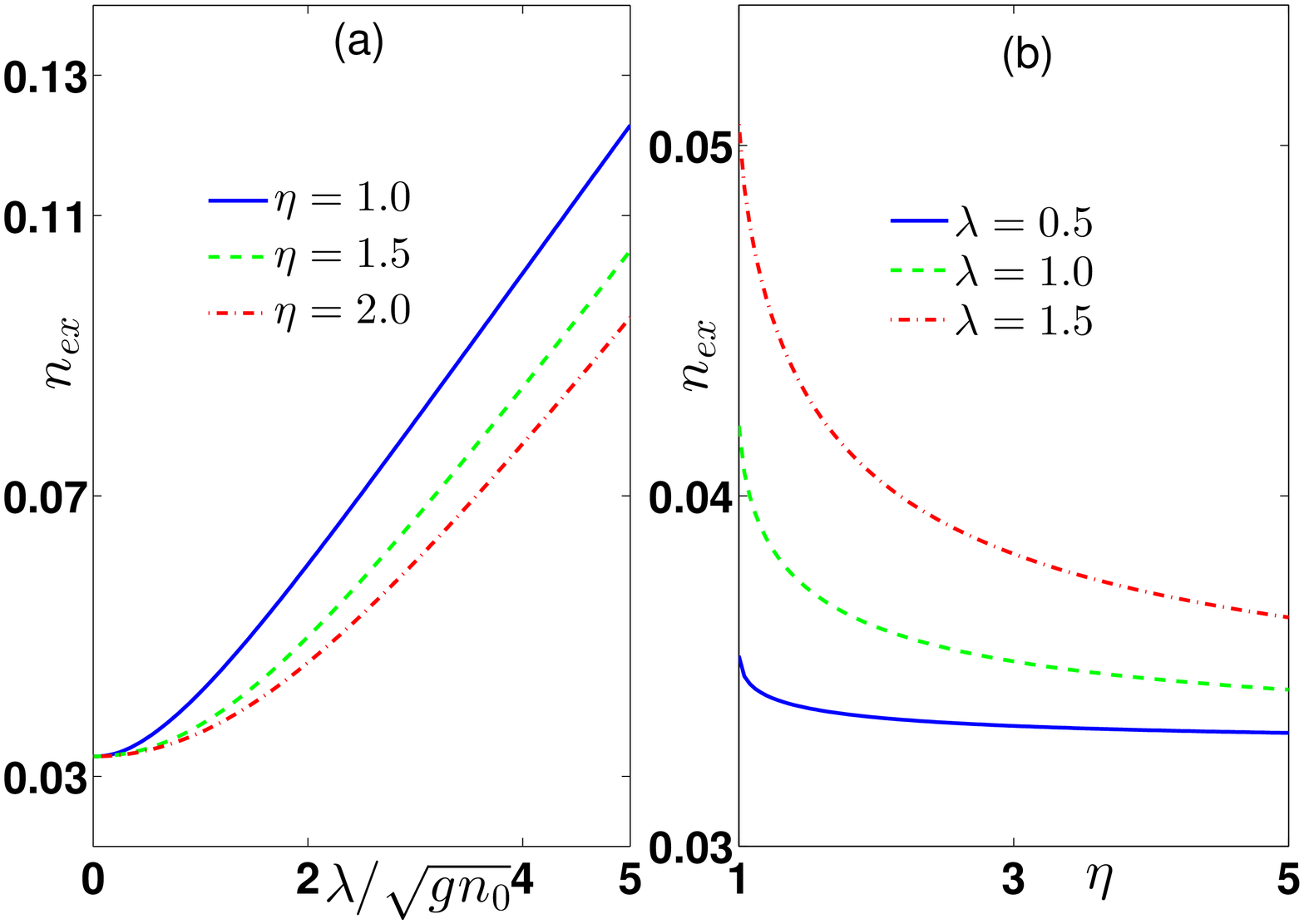}
\caption{(color online) Density of the excited particles due to quantum fluctuation $n_{ex}$ [in units of $(gn_0)^{3/2}$]:  (a) as a function of spin-orbit  coupling strength $\lambda$ for three typical interspecies coupling strength $\eta=1.0$, $\eta=1.5$ and $\eta=2.0$; (b) as a function of interspecies coupling strength $\eta$ for three typical spin-orbit coupling strength $\lambda=0.5\sqrt{gn_0}$, $\lambda=1.0\sqrt{gn_0}$ and $\lambda=1.5\sqrt{gn_0}$.
}
\label{fig4}
\end{figure}

Being an intrinsic property of a BEC, the quantum depletion of the condensates provides vital information concerning the robustness of the superfluid state. The number density of exited particles can be evaluated by employing the quasi-particle's Green's function
\begin{eqnarray}
  n_{ex}=\sum_{\mathbf{q},iw_n}\left[G_{11}(\mathbf{q},iw_n)+G_{22}(\mathbf{q},iw_n)\right].
\end{eqnarray}
We show the density of the excited particles out of the condensates due to quantum fluctuation in Fig.~$\ref{fig4}$. At a fixed interspecies coupling $\eta$, the quantum depletion is monotonically enhanced by spin-orbit coupling, and it reduces to the case of spinless Bose gases with $n_{ex}=(gn_0)^{3/2}/(3\pi^2)$ in the absence of spin-orbit coupling~\cite{UED10}, as seen in Fig.~$\ref{fig4}(a)$.  At a fixed spin-orbit coupling strength, the interspecies coupling actually suppresses quantum depletion, signifying the competing effects of spin-orbit coupling and interspecies coupling upon quantum depletion. When the spin-orbit coupling is small, the effect of interspecies coupling decreases as well, as indicated in Fig.~$\ref{fig4}(b)$. This is quite remarkable, because there is only one species of condensation.  In the absence of spin-orbit coupling, we do not expect that the the interspecies coupling plays any role in quantum depletion. We attribute this behavior to stemming from quantum fluctuation enhanced by spin-orbit coupling.

\begin{figure}[t]
\includegraphics[width=1.0\columnwidth]{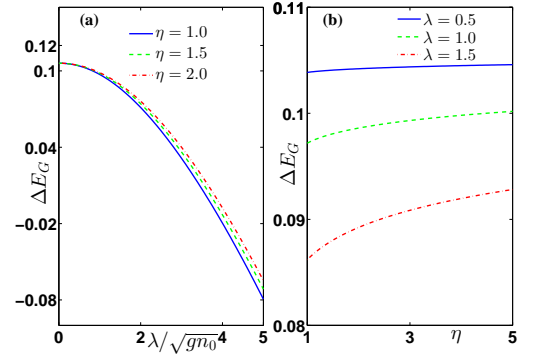}
\caption{(color online) The fluctuation shift of ground state energy  $\Delta E_G=E_G-E_{MF}$ [measured in units of $V(gn_0)^{5/2}$]: (a) as a function of spin-orbit  coupling strength $\lambda$ for three typical interspecies coupling strength $\eta=1.0$, $\eta=1.5$ and $\eta=2.0$; (b) as a function of interspecies coupling strength $\eta$ for three typical spin-orbit coupling strength $\lambda=0.5\sqrt{gn_0}$, $\lambda=1.0\sqrt{gn_0}$ and $\lambda=1.5\sqrt{gn_0}$.
}
\label{fig5}
\end{figure}

The thermodynamic potential of this system is given by $\Omega=-\ln\mathcal{Z}/\beta=\Omega_0+\Omega_f$, where the mean-field part is $\Omega_0=-Vgn_0^2$ and the fluctuating part is $\Omega_f=\frac{1}{2\beta} Trln\mathcal{G}^{-1}-\frac{1}{2}\sum_{\mathbf{q}\sigma} \epsilon_{\mathbf{q}\sigma}$. The thermodynamic potential $\Omega$ possesses an ultraviolet divergence, an artifact of zero range interaction, which can be removed either by replacing the bare interaction g with a $T$ matrix~\cite{STO09} or by subtracting counter-terms~\cite{AND04}. At zero temperature, the ground-state energy becomes $E_G=\Omega+\mu N$, renormalized as
\begin{eqnarray}
  E_G=E_{MF}+\sum_{\mathbf{q}s=\pm}\left[\frac{\omega_{\mathbf{q}s}-(\epsilon_{\mathbf{q}\uparrow}+\epsilon_{\mathbf{q}\downarrow})/2}{2}+\frac{g^2n_0^2}{2q^2}\right].
\end{eqnarray}
Here $E_{MF}=V(gn_0^2-\frac{\lambda^2}{4})$ is the mean-field energy. We show the shift of ground state energy due to quantum fluctuation $\Delta E_G=E_G-E_{MF}$ in Fig.~$\ref{fig5}$. As seen in panel (a), at a fixed  interspecies coupling $\eta$, the shift of the ground state energy $\Delta E_G$ decreases monotonically with the strength of spin-orbit coupling $\lambda$. In the absence of the spin-orbit coupling and interspecies interaction, we have checked that the ground state energy $E_G$ recovers the well-known Lee-Huang-Yang result~\cite{LHY57} for spinless and weakly-interacting Bose gases with $E_G/V=\frac{\mu n}{2}(1+\frac{128}{15\sqrt{\pi}})\sqrt{na^3}$, where $a$ is the scattering length. While, for a finite spin-orbit coupling, the shift of the ground state energy increases with interspecies coupling $\eta$, evidently shown in panel (b).

\begin{figure}[t]
\includegraphics[width=1\columnwidth]{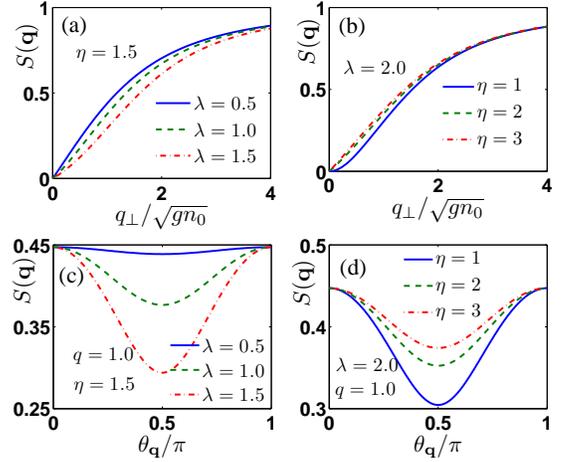}
\caption{(color online) Distribution of the static structure factor $S(\mathbf{q})$ in the momentum space. Upper panel: as a function of in-plane momentum $q_\perp$ for (a) different spin-orbit coupling strength $\lambda$ and (b) different interspecies strength $\eta$. Lower panel: as a function of polar angle $\theta_\mathbf{q}$ for (c) different spin-orbit coupling strength $\lambda$ and (d) different interspecies coupling strength $\eta$.
}
\label{fig6}
\end{figure}
The static structure factor $S(\mathbf{q})$ probes density fluctuations of a system. It provides information on both the spectrum of collective excitations,
which could be investigated at low momentum transfer, and the momentum distribution, which characterizes the behavior of the system at high momentum transfer,
where the response is dominated by single-particle effects. 
At the Bogoliubov level, it can be evaluated as
\begin{eqnarray}
   NS(\mathbf{q})&=&<\delta \rho_\mathbf{q}^\dagger\delta\rho_\mathbf{q}>\nonumber\\
   &=&N_0\sum_{iw_n}\left(G_{11}+G_{33}+G_{13}+G_{31}\right)\nonumber\\
   &=&N_0\sum_{iw_n}\frac{-2q^2A(\mathbf{q},iw_n)}{det[\mathcal{G}^{-1}(\mathbf{q},iw_n)]},
\end{eqnarray}
where $A(\mathbf{q},iw_n)=(iw_n+2\lambda q_z)^2-\omega_{20}^2+\lambda^2\omega_{20}\sin^2{\theta_\mathbf{q}}$. It is quite clear that the static structure factor possesses the cylindrical symmetry $S(\mathbf{q})=S(q,\theta_\mathbf{q})$. At $q_\perp=0$, the static structure factor adopts a close form as follows
\begin{eqnarray}
   S(q_\perp=0,q_z)=\frac{N_0}{N}\frac{q^2}{\omega_{10}}\coth{\frac{\beta\omega_{10}(q)}{2}}.
\end{eqnarray}
In this case, it recovers the Feynman relation~\cite{FEY54,PIT03}, which connects the static structure factor to the excitations spectrum of a Bose system with time-reversal symmetry. We show the behavior of the static structure in Fig.~$\ref{fig6}$. In the upper panel, we show the in-plane static structure factor $S(q,\theta_\mathbf{q}=\pi/2)$ in terms of in-plane momentum $q_\perp$. It decreases as the spin-orbit coupling strength is increased, but increases as the interspecies coupling is increased. Such reversing trend signifies that spin-orbit coupling and interspecies coupling act with reversal role in the density response of the system. In the lower panel, we show angular dependence of the static structure factor at $q=\sqrt{gn_0}$. It is interesting to notice that $S(\mathbf{q})$ is also symmetrical with reflection  about the $x$-$y$ plane, namely $S(q,\theta_\mathbf{q})=S(q,\pi-\theta_\mathbf{q})$.  The static structure factor develops its minimum along $\theta_\mathbf{q}=\pi/2$. The spin-orbit coupling suppresses the density response greatly in the $x$-$y$ plane, as seen in panel (c). In turn, the interspecies coupling enhances the density response greatly along the direction of $\theta_\mathbf{q}=\pi/2$.

\section{Summary and Conclusions}
To sum up, we have studied two-component Bose gases in the presence of Weyl-type SOC. We obtain the phase diagram via a variational approach. We find competing effects between spin-orbit coupling and interspecies coupling strength upon various properties of the PW-Polar phase. There is one crucial difference between them: spin-orbit coupling allows the process of pseudospin flipping process, while interspecies interaction does not permit that. This has far-reaching consequence in the quantum depletion of the condensates. In addition to cylindrical symmetry endorsed by the ground state where the condensation momentum lying along the quantization axis, the sound velocity and the static structure factor also enjoy
a reflection symmetry with respect to $x$-$y$ plane. We hope that our work will contribute to a deeper understanding of  SOC BECs and the role of quantum fluctuations.

\section*{Acknowledgments}
R. L. acknowledges funding from the NSFC under Grants No. 11274064 and NCET-13-0734.

\end{document}